\documentclass[prd,twocolumn,nofootinbib,showpacs,amsmath,amssymb,floatfix,eqsecnum,superscriptaddress]{revtex4-1}

\usepackage{graphicx,color,framed}
\usepackage{hyperref}
\usepackage{times}
\usepackage{enumerate}
\usepackage{lipsum}
\usepackage{slashed}
\usepackage{youngtab}

\hypersetup{
    colorlinks=true, 
    linktoc=all,     
    linkcolor=blue,  
}

\def \beq {\begin{equation}}
\def \eeq {\end{equation}}
\def \beqa {\begin{eqnarray}}
\def \eeqa {\end{eqnarray}}
\def \bseq {\begin{subequations}}
\def \eseq {\end{subequations}}
\newcommand \dg {\dagger}
\newcommand \al {\alpha}
\newcommand \be {\beta}

\newcommand \ran {\rangle}
\newcommand \lan {\langle}

\newcommand \ep {\epsilon}

\newcommand \pd {\partial}

\newcommand \vphi {\varphi}

\newcommand \ov {\overline}

\def \be {\begin{eqnarray}}
\def \ee {\end{eqnarray}}
\newcommand \nn{\nonumber}
\newcommand{\eqn}[1]{(\ref{#1})}
\newcommand{\ttt}[1]{\text{\tiny{#1}}}

\begin{document}

\title{Hall Viscosity in the Non-Abelian Quantum Hall Matrix Model}

\author{Matthew F. Lapa}
\email{lapa2@illinois.edu}
\affiliation{Department of Physics and Institute for Condensed Matter Theory, University of Illinois at Urbana-Champaign, Urbana, IL, 61801-3080, USA}
\author{Carl Turner}
\email{c.p.turner@damtp.cam.ac.uk}
\affiliation{Department of Applied Mathematics and Theoretical Physics, University of Cambridge, Cambridge, CB3 0WA, UK}
\author{Taylor L. Hughes}
\email{hughest@illinois.edu}
\affiliation{Department of Physics and Institute for Condensed Matter Theory, University of Illinois at Urbana-Champaign, Urbana, IL, 61801-3080, USA}
\author{David Tong}
\email{d.tong@damtp.cam.ac.uk}
\affiliation{Department of Applied Mathematics and Theoretical Physics, University of Cambridge, Cambridge, CB3 0WA, UK}




\begin{abstract}

Quantum Hall matrix models are simple, solvable quantum mechanical systems which capture the physics of certain fractional quantum Hall states. 
Recently, it was shown that the Hall viscosity can be extracted from the matrix model for Laughlin states. Here we extend this calculation to the 
matrix models for a class of non-Abelian quantum Hall states. These states, which were previously introduced by Blok and Wen, arise from the conformal blocks of Wess-Zumino-Witten conformal field theory models. We show that the Hall viscosity computed from the matrix model coincides with a result of Read, in which the Hall viscosity is determined in terms of the weights of primary operators of an associated conformal field theory.

\end{abstract}


\maketitle

\section{Introduction}

Quantum Hall matrix models provide examples of solvable many-body quantum systems. The ground states of these theories coincide with well-known quantum Hall states, both Abelian \cite{P1} and non-Abelian \cite{tong2016}. Meanwhile, the partition function of these models reproduces the edge dynamics of the corresponding boundary chiral conformal field theory (CFT) \cite{MM-WZW}.

Recently, it was shown how one can extract the transport coefficient known as the Hall viscosity from the matrix model for the Laughlin states \cite{lapa2018}. The purpose of this article is to extend this computation to the matrix models describing non-Abelian quantum Hall states. As we will see, this makes contact with a general result due to Read, in which the Hall viscosity is related to the dimension of a certain primary operator in the boundary CFT \cite{read2009}. Furthermore this provides a more non-trivial connection between the matrix models and the geometric response properties of fractional quantum Hall states.

In the remainder of this introduction, we describe some of the necessary background material on Hall viscosity. 
We will introduce the relevant matrix models in Section \ref{mmsec}, and the calculation of the Hall viscosity for the non-Abelian quantum Hall states can be found in Section \ref{mmhvsec}. We note here that in Section \ref{mmhvsec} we give 
only a brief review of how the Hall viscosity can be extracted from the matrix model. Readers interested in 
the full details of this procedure should consult Ref.~\onlinecite{lapa2018}.

\subsection{Review of Hall Viscosity}

The Hall viscosity $\eta_H$, first introduced in \cite{ASZ}, characterises the response of a state to changes in the geometry, and has been the focus of much subsequent interest \cite{levay,avron1998odd,TV1,read2009,TV2,haldane2009,haldane2011,read-rezayi,HLF2011,hoyos-son,bradlyn2012,park-haldane}. In addition, 
it is now known that Hall viscosity is just one 
aspect of a larger program of study focusing on the geometric properties of fractional quantum Hall 
states~\cite{AG,cho2014,YCF-nematic,ferrari-klevtsov,BR1,BR2,CLW,framing,KN,GGB,gromov-son}.
In generic fractional quantum Hall (FQH) states, the Hall viscosity is quantised as~\cite{read2009}
\be \eta_H = \frac{\hbar{\cal S}}{4}\rho_0,\nn\ee
where $\rho_0$ is the average density of the Hall state, given by $\rho_0 = \frac{\nu}{2\pi\ell_B^2}$ where $\nu$ the filling fraction, and $\ell_B^2=\frac{\hbar}{eB}$ is the square of the magnetic length (here $B$ is the magnetic field and $-e<0$ is the electronic charge). The quantised nature of the Hall viscosity lies in the fact that ${\cal S} \in \mathbb{Z}$; this integer is referred to as the {\it shift}, and it represents an order one correction to the usual relation $N_{\phi}=\nu^{-1}N$ between $N$, 
the number of electrons in a FQH state, and $N_{\phi}$, the number of flux quanta of the magnetic field piercing the 
sample, when the FQH state is placed on a sphere. Indeed, on the sphere one has~\cite{WZ}
\beq
	N_{\phi}=\nu^{-1}N - \mathcal{S}\ .\nn
\eeq
For the Laughlin states, with $\nu = 1/k$, the shift is given by ${\cal S} = k$. This means that the Hall viscosity is actually independent of the level $k$ for Laughlin states, since ${\cal S}\rho_0 = 1/(2\pi\ell_B^2)$.

Some years ago, Haldane suggested that there is more to this formula than meets the eye \cite{haldane2009,haldane2011}. He argued that the Hall viscosity should be viewed as the sum of two terms
\be \eta_H = \eta_{\text{\tiny{LO}}} + \eta_{\text{\tiny{GC}}}\nn,\ee
referred to as the {\it Landau orbit} and {\it guiding center} contributions respectively.  The Landau orbit contribution comes from the cyclotron motion of the electrons in the external magnetic field. For states in the lowest Landau level, it is given by  
\be \eta_{\ttt{LO}} = \frac{\hbar\rho_0}{4}\ .\label{etalo}
\ee
Note that the Landau orbit contribution is non-vanishing even for integer quantum Hall states. The guiding center contribution is more interesting; it is non-vanishing only in fractional quantum Hall states, and can be related to the electric dipole moment per unit length along the edge of the state~\cite{park-haldane}.  Each of these contributions (Landau orbit and guiding center) is individually quantised and, correspondingly, the shift can be decomposed as~\cite{GGB,gromov-son}
\be {\cal S} = 2(s+\varsigma)\label{shiftdecomp}\ .\ee 
Here $s\in \frac{1}{2}\mathbb{Z}$  is associated to the Landau orbit contribution. 
Meanwhile,  $\varsigma \in \frac{1}{2}\mathbb{Z}$ (an alternative  ``sigma" symbol) is the {\it anisospin}, and is associated to the guiding center contribution\footnote{The terminology \emph{anisospin} was introduced in \cite{GGB,gromov-son}, and refers to the fact that $\varsigma$ captures the response of the system to anisotropy. 
This coefficient is  related to the \emph{guiding center spin} $s_{\text{{\tiny{GC}}}}$ defined by 
Haldane~\cite{haldane2009,haldane2011} as $s_{\text{{\tiny{GC}}}}=-\varsigma$. A useful table
listing the values of $\varsigma$ in several different families of FQH states can be found in 
Appendix A of \cite{gromov-son}.}. Note that the Laughlin state with $\nu = 1/k$ has $\varsigma = (k-1)/2$, and so the guiding center Hall viscosity does distinguish between different states.

Projecting to the lowest Landau level yields only the guiding center contribution to the Hall viscosity. This fact will be important for us, so we offer a fuller explanation.  For a trial quantum Hall wave function, the total Hall viscosity can be computed using the generators of 
\emph{area-preserving deformations} of the full electron coordinates~\cite{haldane2009,haldane2011,park-haldane}.
Those generators break up into a sum of a guiding center part and a Landau orbit part, and these two parts give rise to the
guiding center and Landau orbit contributions to the Hall viscosity~\cite{haldane2009,haldane2011,park-haldane}. 
However, if we first project into the lowest Landau level
and then calculate the Hall viscosity, we would do the calculation using the projected area-preserving deformation generators,
and the projected generators would consist only of the guiding center part. As a result, the Hall viscosity computed after
the projection would contain only the guiding center part. We will see this explicitly in the matrix model computations later in the paper.

The low-energy effective description of  quantum Hall states offers a useful perspective on the story above. The simplest such description is a topological field theory. For the Laughlin states,  with $\nu = 1/k$, this theory involves an emergent dynamical $U(1)$ gauge field $a$, coupled to the background electromagnetic gauge field $A$. To capture the response to the geometry, we also introduce a background $O(2)$ spin connection $\omega$. The low-energy dynamics is then described by the Chern-Simons theory 
\be S = \frac{\hbar k}{4\pi}\int ada -\frac{e}{2\pi}\int Ada - \frac{\hbar{\cal S}}{4\pi} \int ad\omega\label{tft}\ .\ee
More recently, it was suggested that there is a more general effective field theory, in which one moves beyond the strictly topological sector \cite{GGB,gromov-son}. In this framework, the decomposition of the shift \eqn{shiftdecomp} is particularly natural. One introduces a second, dynamical spin connection $\hat{\omega}$, whose role is to describe the lowest energy, gapped spin 2 density perturbations seen in quantum Hall states, often referred to as the magnetoroton, or
Girvin-MacDonald-Platzman (GMP) mode \cite{gmp}. This new ``bi-metric" theory contains two mixed gauge-gravitational Chern-Simons terms, with quantised coefficients
\be S = \frac{\hbar k}{4\pi}\int ada -\frac{e}{2\pi}\int Ada - \frac{\hbar s}{2\pi} \int ad\omega - \frac{\hbar\varsigma}{2\pi} \int ad\hat{\omega} + \ldots \nn\ee
where $\ldots$ includes dynamical information about the propagating mode. Note that the Gauss-Bonnet theorem
for the spin connection $\omega$ (and also $\hat{\omega}$) implies that the coefficients $s$ and $\varsigma$ are
quantised as $s,\varsigma \in \frac{1}{2}\mathbb{Z}$, as opposed to being restricted to integer values.\footnote{In more detail, if 
$\mathcal{M}$
is the two-dimensional space and $\omega$ is the spin connection for $\mathcal{M}$, then by Gauss-Bonnet, 
$\int_{\mathcal{M}}d\omega= 2\pi \chi$, where $\chi$ is the Euler characteristic of $\mathcal{M}$ and is an \emph{even}
integer for orientable $\mathcal{M}$. The main point is that $\int_{\mathcal{M}}d\omega$ is an even integer multiple of $2\pi$, as opposed to just an integer multiple of $2\pi$ (which is the case for ordinary $U(1)$ gauge fields obeying the Dirac quantization condition).} At long wavelengths, the equation of motion requires $\hat{\omega} = \omega$, and the dynamical theory above reduces to the purely topological theory \eqn{tft}, with the shift given by \eqn{shiftdecomp}.


\subsection*{Hall Viscosity: The View from the Boundary}

It has long been known that the edge modes of the quantum Hall state carry a surprising amount of information about the bulk. 
The key idea dates back to Moore and Read \cite{MR}, who showed that the bulk quantum Hall wavefunctions could be reinterpreted as conformal blocks of the boundary CFT.  Specifically, one can construct a trial quantum Hall wavefunction using the ansatz
\beq\begin{split}
	\Psi_{\text{trial}}(z_1,\dots,z_N)&= \Psi_{\text{charge}}(z_1,\dots,z_N)
	\\
	&\qquad \times \lan \mathcal{O}(z_1)\cdots\mathcal{O}(z_N)\ran_{\text{CFT}}\ , \label{eq:trial}
\end{split}\eeq
where $z_1,\dots,z_N$ are the (holomorphic) electron coordinates. The first factor,  $\Psi_{\text{charge}}$, is a correlation function of a free, compact boson; this gives a Laughlin-like contribution to the wavefunction and determines the filling fraction $\nu$. (If the action for the free boson is normalized to have a coefficient of $1/8\pi$, then the filling fraction is given by
$\nu = R^{-2}$, where  $R$ is the radius of the boson.)
The second factor, 
$\lan \mathcal{O}(z_1)\cdots\mathcal{O}(z_N)\ran_{\text{CFT}}$, is a correlation function in a different CFT and determines other properties of the Hall state, including any non-Abelian statistics.

The electron creation operator on the boundary is identified with ${\cal O}(z)$, weighted with a vertex operator from the compact boson. 
The operator $\mathcal{O}(z)$ is a primary field in the CFT. It must also satisfy a more technical property: it must be a {\it simple current} which, roughly speaking, means there is a unique fusion channel when it is fused any other primary operator.


%

In \cite{read2009}, Read studied quantum Hall states of the form \eqn{eq:trial} defined on a torus. By adiabatically varying the modular parameter (i.e. the aspect ratio) of the torus, he was able to determine the Hall viscosity in terms of the boundary CFT data; it is given by
\beq
	\eta_{H}= \frac{\hbar}{2}\left( \frac{1}{2\nu} + h_{\mathcal{O}}\right) \rho_0\ ,\label{readtotal}
\eeq 
where $h_{\mathcal{O}}$ is the conformal weight of the primary field $\mathcal{O}(z)$. Since the state \eqn{eq:trial} lies in the lowest Landau level, the generalisation of Read's result to the guiding center Hall viscosity is straightforward: we simply subtract off the constant Landau orbit contribution \eqn{etalo}. We are left with
\beq
	\eta_{\text{{\tiny{GC}}}}= \frac{\hbar}{2}\left( \frac{1}{2\nu} -\frac{1}{2} + h_{\mathcal{O}}\right) \rho_0\label{read}\ .
\eeq
We will make contact with this formula later in the paper. 

\vskip 6mm

\section{A Matrix Model for Non-Abelian Quantum Hall States}\label{mmsec}

In this section we describe a matrix model for non-Abelian quantum Hall states. As we will see, our particular matrix model describes a class of states first introduced by Blok and Wen \cite{blok-wen}. These are the Hall states that are described by $U(p)_k$ bulk Chern-Simons theory or, equivalently, by $U(p)_k$ Wess-Zumino-Witten CFTs on the boundary.

\subsection{Overview of the Matrix Model}

The matrix model is quantum mechanical system, comprising a pair of  $N\times N$ Hermitian matrices $X^a(t)$, $a=1,2$, 
and a set of $p$ vectors $\vphi_{\al}(t)$, $\al=1,\dots,p$, each of dimension $N$.  These are coupled through one further Hermitian matrix,  $A_0(t)$, which should be viewed as a $U(N)$ gauge field. The action is given by
\begin{eqnarray}
	S_{\text{{\tiny{MM}}}} = -\int dt\  \text{Tr}\Big\{ &&\frac{eB}{2}\ep_{ab}X^a {\cal D}_0 X^b + \hbar (k+p)A_0 \label{mmact}\\
&&+\ \frac{eB\omega}{2} \delta_{ab}X^a X^b \Big\} + \sum_{\al=1}^p\int dt\ i\ov{\vphi}^{T}_{\al} {\cal D}_0{\vphi}_{\al}\ . \nn
\end{eqnarray}
Here the covariant time derivatives are defined by\footnote{A comment on notation: we use $[\cdot,\cdot]_M$ 
to denote a commutator of classical matrix variables, and $[\cdot,\cdot]$, with no subscript, for a commutator of quantum operators. Relatedly, $ \ov{\vphi}{}^{T}_{\al}$ denotes the
row vector whose components are the complex conjugates of the components of the column vector $\vphi_{\al}$. We reserve $\dg$ to denote Hermitian conjugation of quantum operators.}
\beqa
	{\cal D}_0 X^b &=& \dot{X}^b - i[A_0,X^b]_M  \\
  {\cal D}_0\varphi &=& \dot{\varphi} - i A_0\varphi \ .
\eeqa
The action \eqn{mmact} describes the dynamics of $N$ particles, each of charge $-e$ (with $e>0$), moving in a magnetic field $B$ and projected to the lowest Landau level. The positions of these particles are, roughly speaking, to be thought of as the eigenvalues of the matrices $X^a$. (Here, the ``roughly speaking" is because these matrices do not commute with each other, and so are not simultaneously diagonalisable.) The idea that the dynamics of $N$ particles can be described by an $N\times N$ matrix is familiar from D-brane physics \cite{witten} and, indeed, from random matrix theory. However, it may be less familiar in a condensed matter context. We will elaborate on this interpretation as we go along.

The matrix model with  $p=1$ was first introduced by Polychronakos in \cite{P1} and, as we review below,  reproduces the Laughlin states at filling fraction $\nu=1/(k+1)$. The matrix model with $p>1$ has an $SU(p)$ global symmetry and was shown to describe non-Abelian quantum Hall states in \cite{tong2016}. (This model was previously discussed in \cite{MP}, albeit with a somewhat different interpretation.)

The action for the matrix model enjoys a  $U(N)$ gauge symmetry, under which the fields transform as
\beqa
	X^a \to VX^a \ov{V}{}^T\ \ \text{and}\ \ \vphi_{\al} \to V\vphi_{\al}\ ,\  \ \  \label{gaugetrans}
\eeqa
and $A_0 \to VA_0 \ov{V}^T + i V \dot{\ov{V}}{}^T$, where $V(t)$ is a time-dependent $U(N)$ matrix (and $\ov{V}{}^T$ is its transpose conjugate). More precisely, the action is usually invariant under such gauge transformations, but not always. The culprit is the quantum mechanical Chern-Simons term, proportional to $\text{Tr}\,A_0$. Suppose that we compactify the time direction, so that $t\in [0,T)$, and impose periodic boundary conditions. Then the action is not invariant under large gauge transformations, in which $V(t)$ winds around the temporal circle, using fact that $\Pi_1(U(N)) = \mathbb{Z}$. This, of course, is a standard story: the action $S_{\text{\tiny{MM}}}$ is not invariant, but the quantum theory depends only on $e^{iS_{\text{{\tiny{MM}}}}/\hbar}$, and this is invariant provided that the coefficient of the Chern-Simons term is quantised. In the present case, this requires
\be k+p\ \in\ \mathbb{Z}\ .\nn\ee
Since $p\in \mathbb{Z}$, we clearly must have $k\in \mathbb{Z}$. In what follows, we take $k> 0$.  The same condition also arises in the canonical quantisation approach, as we will see below.

Before proceeding, we pause to explain a little more about the origin of the matrix model. The original motivation (for the $p=1$ matrix model) came from work of Susskind, who proposed a hydrodynamic description of the quantum Hall fluid in terms of Chern-Simons theory on a non-commutative plane \cite{susskind}. This can be viewed as the $N\rightarrow \infty$ limit of Polychronakos's matrix model, in which the $\varphi$ field becomes redundant, and the matrices $X^a$ become coordinates on the non-commutative plane.

Subsequently, an alternative interpretation of the matrix model was given in \cite{oldstory,susyqhe}, where it was shown that it describes the dynamics of $N$ vortices in (ordinary) $U(p)$ Chern-Simons theory, coupled to non-relativistic scalar fields.  To see this connection, we first elucidate some further structure of the matrix model. The gauge field $A_0$ acts as a Lagrange multiplier, enforcing the Gauss law constraint,
\beq
	G\ \equiv\  ieB[X^1,X^2]_M+ \hbar (k+p) \mathbb{I} - \sum_{\alpha=1}^p \vphi_{\al}\ov{\vphi}^T_{\al}= 0\ , \label{eq:constraint}
\eeq
where $\mathbb{I}$ is the $N\times N$ identity matrix. Working in the $A_0=0$ gauge, 
physically distinct field configurations consist of matrices $X^a$ and vectors $\varphi_\alpha$, subject to the constraint \eqn{eq:constraint}, and the gauge quotient \eqn{gaugetrans}. Solutions to these equations can be thought of as parametrising a manifold ${\cal M}$ of dimension ${\rm dim}({\cal M}) = 2Np$. Because the original action \eqn{mmact} is first order in time derivatives, rather than second order, ${\cal M}$ is the phase space of our system.

Yet this same manifold ${\cal M}$, defined by equations \eqn{gaugetrans} and \eqn{eq:constraint}, has appeared elsewhere: these equations are known to describe the configuration space (usually referred to as {\it moduli space}) of BPS vortices in $U(p)$ gauge theories. This statement was originally shown using D-brane techniques \cite{ht}, and subsequently received a more mathematically rigorous treatment in \cite{nakajima}. Since the matrix model treats ${\cal M}$ as a phase space, rather than configuration space, it can be viewed as enacting a geometric quantisation of the moduli space of vortices.

Finally, we are left with the Hamiltonian of the matrix model. This is simply
\beq
	H_{\text{{\tiny{MM}}}} = \frac{eB\omega}{2}\,\text{Tr}\{\delta_{ab}X^a X^b\}\ .
\eeq
This has the interpretation of a harmonic trap, with strength determined by the frequency $\omega$, 
encouraging the fluid of particles to cluster near the origin.

\subsection{Quantization and Ground State}

We now give a brief discussion of the quantization of the matrix model and the construction of the quantum ground state.
The quantization of the matrix model was first described in \cite{P1,HVR}. Each of the components of the matrices $X^a,$ and vectors $\varphi_\alpha,$ now become operators on a quantum Hilbert space.  We use lowercase Latin indices $j,k,\ell,\dots$ from the middle of the alphabet\footnote{The index $k$ should not be confused with the level of the Chern-Simons term in \eqn{mmact}.} to label
the matrix elements  of $X^a$ and the components of $\vphi_{\al}$, for example ${(X^a)^j}_k$ and $\vphi_{\al}^j$ (with 
$j,k=1,\dots,N$). 



As in \cite{lapa2018}, we parametrize the matrices $X^a(t)$ in terms of a set of $N^2$ real scalar
variables $x^a_A(t)$, $A=0,\dots,N^2-1$, 
by expanding $X^a(t)$ using a set of generators $T^A$ of the Lie algebra  $su(N)$,
\beq
	X^a(t)= x^a_0(t)\frac{\mathbb{I}}{\sqrt{N}} + \sum_{A=1}^{N^2-1} x^a_A(t) T^A\ . \label{eq:parametrization}
\nn\eeq
We choose to normalize\footnote{In the fundamental representation the correct normalization is 
$\text{tr}\{T^AT^B\} = \frac{1}{2}\delta^{AB}$. However, since we are only using these generators as a basis for the space 
of $N\times N$ matrices, we are free
to choose a normalization which is convenient for this purpose.}  the generators as $\text{tr}\{T^A T^B\}= \delta^{AB}$. For later convenience
we also define $T^0:=\frac{\mathbb{I}}{\sqrt{N}}$. With this definition the relation $\text{tr}\{T^A T^B\}= \delta^{AB}$
holds for all values of $A,B$ including $0$. In the quantized theory the $x^a_A$ variables obey the commutation
relations
\beq
	[x^1_A,x^2_B]= i\ell_B^2\delta_{AB}\ . \label{eq:x-relations}
\eeq
We further define 
\beq
	z_A = \frac{1}{\ell_B\sqrt{2}}( x^1_A + i x^2_A)\ \ \ {\rm and}\ \ \ b^j_{\al}= \frac{1}{\sqrt{\hbar}}\vphi^j_{\al} \ ,
\eeq
where $\ell_B^2 = \hbar/(eB)$ is the usual magnetic length. The $z_A$ provide complex coordinates on the plane.
 Correspondingly, we write $z^{\dg}_A = \frac{1}{\ell_B\sqrt{2}}( x^1_A - i x^2_A)$, and $b^{\dg}_{\al,j}= \frac{1}{\sqrt{\hbar}}\ov{\vphi}_{\al,j}$. For later use we also define the matrix $Z^{\dg}$ whose components ${(Z^{\dg})^j}_{k}$ are quantum operators, 
\beq
	{(Z^{\dg})^j}_{k}=  \sum_{A=0}^{N^2-1} z^{\dg}_A {(T^A)^j}_{k}\ , \label{eq:matrix-op}
\nn\eeq
where we remind the reader that $T^0:=\mathbb{I}/\sqrt{N}$.

Upon quantization, the Poisson brackets of \eqn{mmact} turn into commutators in the usual fashion. The resulting commutation relations are
\beqa
	\left[z_A, z^{\dg}_B\right] &=& \delta_{AB} \nn\\
	\left[ b^j_{\al}, b^{\dg}_{\beta,\ell}\right]&=& \delta_{\al\beta}\delta^j_{\ell}\nn\ ,
\eeqa
together with $[z_A,b_\alpha^j]=[z_A,b_{\alpha,\ell}^\dg]=0$. These, of course, are simply a collection of harmonic oscillator creation and annihilation operators. The Fock vacuum, $|0\ran$, is defined to obey $z_A|0\ran = b_{\al}^j|0\ran=0$. All other states in the Fock space are built by acting with creation operators $z_A^\dg$ and $b^\dg_{\alpha,\ell}$. The complication is that not all these states are physical; any physical state $|\psi\ran$ must obey the quantum version of the Gauss' law constraint \eqn{eq:constraint}:
\beq
	G^A|\psi\ran = 0\ \ \ \ {\rm for}\ A=0,\ldots, N^2-1\ .
\label{qconstraint}\eeq
Here $G^A$ are obtained from the contraction of $G$ from Eq.~\ref{eq:constraint} with the generators $T^A$ as $G^A = {G^j}_k {(T^A)^k}_j$.
As explained in \cite{P1}, the equations \eqn{qconstraint} have two distinct components. The traceless part (corresponding to $A=1,\ldots N^2-1$) requires that physical 
states are singlets under the $SU(N) \subset U(N)$ gauge group. On the other hand, the trace of the constraint (corresponding to $A=0$) tells us that physical states carry a specific charge under $U(1)\subset U(N)$, 
\beq
	\sum_{\al=1}^p b_{\al}^j b^{\dg}_{\al,j} |\psi\ran= N(k+p)|\psi\ran\ .
\nn\eeq
This is the role played by the Chern-Simons term in the matrix model.  After exchanging the order of $b_{\al}^j$ and 
$b^{\dg}_{\al,j}$ using their commutation relation, this becomes
\beq
	\sum_{\al=1}^p b^{\dg}_{\al,j}b_{\al}^j  |\psi\ran= Nk|\psi\ran\ . \label{eq:U1-constraint}
\eeq
We learn that all physical states must carry charge  $Nk$ under the $U(1)\subset U(N)$. Note that the left-hand side is a number operator and so clearly an integer. The same must be true of the right-hand side. In this way, we see again that the matrix model requires $k\in \mathbb{Z}$.

The Hamiltonian for the matrix model (in the $A_0=0$ gauge), takes the form
\begin{eqnarray}
	H_{\text{{\tiny{MM}}}} &=& \frac{eB\omega}{2}\sum_{A=0}^{N^2-1}\delta_{ab}x^a_A x^b_A \nn \\ & =& \hbar\omega\left(\frac{N^2}{2} + \sum_{A=0}^{N^2-1} z^{\dg}_A z_A \right)\ .
\end{eqnarray}
We recognise the first term as the zero-point energy of $N^2$ harmonic oscillators. The second term is the number of $z_A$
 quanta contained in a state.

This, then, is our task: to construct the physical Hilbert space we must form $SU(N)$ singlets with exactly $Nk$ excited $b^\dg_{\alpha,j}$ quanta. The energy of these states is proportional to the number of the excited $z_A^\dg$ quanta.

The  physical ground state of the matrix model is the physical state with the fewest $z_A^\dg$ excitations. For $p>1$, this was constructed in  \cite{tong2016}, and follows closely the earlier work in \cite{P1,HVR}.  First, for any integer $r\geq 0$ we construct what a particle physicist would refer to as a ``baryon operator", 
\beq
 \mathcal{B}^{\dg}(r)_{j_1\cdots j_p} := \ep^{\al_1\cdots\al_p}[b^{\dg}_{\al_1}(Z^{\dg})^r]_{j_1}\cdots[b^{\dg}_{\al_p}(Z^{\dg})^r]_{j_p}\ .
\nn\eeq
This operator is, by construction, a singlet under the $SU(p)$ global symmetry of the matrix model, but transforms in the  $p^{\text{th}}$ 
antisymmetric representation of the  $SU(N)$ gauge symmetry. To proceed, life is simplest if we assume that $N$ is divisible by $p$. (See \cite{tong2016} for the more general case.) We can then construct a ``baryon of baryons", which is a singlet under both $SU(p)$ and $SU(N)$,
\beq\begin{split}
\tilde{\mathcal{B}} &:= \ep^{j_1\cdots j_N}\mathcal{B}^{\dg}(0)_{j_1\cdots j_p}\mathcal{B}^{\dg}(1)_{j_{p+1}\cdots j_{2p}}  \\ &\hspace{6em}\cdots\mathcal{B}^{\dg}({N}/{p}-1)_{j_{N-p+1}\cdots j_N}\ .
\nn\end{split}\eeq
This operator is the $SU(N)$ singlet with the fewest $z_A^\dagger$ excitations. It carries charge $N$ under the $U(1) \subset U(N)$ gauge symmetry. Since physical states must carry charge $kN$, we learn that the unique ground state of the matrix model is then given by 
\beq
	|\psi_0\ran = \tilde{\mathcal{B}}^k|0\ran \ .\label{ground}
\eeq

\subsection*{The Filling Fraction}

The quantum Hall states \eqn{ground} are written in an abstract form. However, as we describe in more detail in Section \ref{bwsec}, it turns out that these are well known quantum Hall states.  First we explain how to compute  the filling fraction of the states \eqn{ground}.

The filling fraction can be determined by the radius of the disc of Hall fluid which, in turn, is determined by the angular momentum of the state. The angular momentum operator in the matrix model is straightforward to identify: it is the Noether charge associated to rotations $Z\rightarrow e^{i\alpha}Z$ ($Z\sim X^1+iX^2$), given by
\begin{eqnarray}
	L &=& -\frac{eB}{2}\sum_{A=0}^{N^2-1}\delta_{ab}\,x^a_A x^b_A \nn\\ &=&  -\hbar\left(\frac{N^2}{2}+ \sum_{A=0}^{N^2-1}z^{\dg}_A z_A \right)\ .
\end{eqnarray}
To compute the angular momentum in the ground state $|\psi_0\ran$ we note that each baryon operator $\mathcal{B}^{\dg}(r)_{j_1\cdots j_p}$ contains $p r$ $z_A$ quanta. The total  number of $z_A$ quanta in the state $|\psi_0\ran$ is then
\beq
	kp\sum_{r=0}^{N/p-1} r= \frac{kN}{2}\left(\frac{N}{p}-1\right)\ . \nn
\eeq
The angular momentum, $L_0$, of the ground state \eqn{ground} is then given by 
\beqa
	L_0 &=& -\hbar\left[\frac{N^2}{2}+\frac{kN}{2}N\left(\frac{N}{p}-1 \right) \right] \nn\\ & =&  -\hbar\left(\frac{k+p}{p}\right)\frac{N^2}{2} + \hbar\frac{kN}{2}\ .\label{l0}
\eeqa
The filling fraction can be read off from the coefficient of the leading $N^2$ term; it is 
\beq
	\nu= \frac{p}{k+p}\label{ffraction}\ .
\eeq
This gives the mean density of the state, 
\beq
	\rho_0=\frac{\nu}{2\pi\ell_B^2}= \left( \frac{p}{k+p}\right)\frac{1}{2\pi \ell_B^2}\label{density}\ .
\eeq
Note that since $\rho_0={N}/{\mathcal{A}}$, the  area of the quantum Hall droplet is
\beq
	\mathcal{A}= 2\pi\ell_B^2\left(\frac{k+p}{p}\right)N\ . \label{eq:MM-area}
\eeq
In particular, the area is extensive (i.e., linear in $N$).

\subsection{Comparison to Blok-Wen States}\label{bwsec}

The abstract states \eqn{ground} represent well known quantum Hall states.  For $p=1$, they are simply the Laughlin states, with filling fraction $\nu=1/(k+1)$. Meanwhile, for $p>1$, they are a class  of states  first constructed by Blok and Wen \cite{blok-wen} with filling fraction \eqn{ffraction}. 
In this section, we give a brief description of how these states are identified and, for the Blok-Wen states, some of their defining properties. 

Let us start with the Laughlin states that arise in the matrix model with $p=1$. Here, the ground state \eqn{ground} can be written as
\be |\psi_0\ran = \left(\epsilon^{j_1\ldots j_N}b^\dg_{j_1}\,[b^\dg Z^\dg]_{j_2}\,[b^\dg (Z^{\dg})^2]_{j_3}\ldots[\,b^\dagger (Z^{\dg})^{N-1}]_{j_N}\right)^k|0\ran\nn\ee
which has the same formal structure as the power of the Vandermonde determinant
that appears in the Laughlin state\footnote{The power of the Vandermonde 
determinant which appears in the $\nu=1/m$ Laughlin state can be rewritten as
$\prod_{1\leq j < k \leq N}(z_j-z_k)^m= \left[-\ep^{j_1\cdots j_N}z_{j_1}^0 z_{j_2}^1\cdots z_{j_N}^{N-1}\right]^m$.}. 
To make the connection with the Laughlin state more precise, one can study the wave function 
$\psi_0(Z) = \lan Z | \psi_0\ran$ obtained by taking the inner product of $|\psi_0\ran$ with a coherent state
$|Z\ran$ for the operator $\hat{Z}$, as first explained by 
Karabali and Sakita \cite{ks1,ks2}. The 
coherent state $|Z\ran$ is defined by\footnote{Here 
$\hat{Z}^j{}_k=\sum_{A=0}^{N^2-1}\hat{z}_A {(T^A)^j}_k$. In this 
paragraph only, we use a hat to distinguish quantum operators from c-numbers.} 
$\hat{Z}^j{}_k|Z\ran= {Z^j}_k |Z\ran$. After obtaining $\psi_0(Z)$ in this way, one
can go further and express this wave function in terms of only the eigenvalues $z_j$, $j=1,\dots,N,$ of the complex
matrix $Z$, so that $\psi_0(Z)\to\psi_0(z_1,\dots,z_N)$. The net result of this analysis is that, at large distances, so that the 
separation between particles is greater than 
$\ell_B$, the wave function $\psi_0(z_1,\dots,z_N)$ coincides with the Laughlin wave function at filling fraction $\nu = 1/(k+1)$. 
(The shift from $k$  to $k+1$ can be traced to the Jacobian factor which arises in moving from the matrix $Z$ to its 
eigenvalues.)

For  $p>1$, the story is similar, but comes with an important extra ingredient: each particle carries an internal ``spin" degree of freedom, transforming in a particular representation of $SU(p)$. To see this, we need only look at the matrix model for  $N=1$, describing a single particle. The ground state \eqn{ground} is no longer unique (since $N=1$ is not divisible by $p$) but instead takes the form
\be |\psi_0\ran_{\alpha_1\ldots \alpha_k} = b^\dagger_{\alpha_1}\ldots b^\dg_{\alpha_k}|0\ran\ .\label{eq:single-particle-states}\ee
This lack of uniqueness reflects the internal degree of freedom of the particle. Since each $b_\alpha$ transforms in the fundamental of $SU(p)$, the single particle ground state transforms in the $k^\text{th}$ symmetric representation of $SU(p)$.

The multi-particle states \eqn{ground} can then be viewed as spin singlet quantum Hall states, where each particle transforms in the $k^\text{th}$ symmetric representation of $SU(p)$. Explicit forms of the resulting wavefunctions can be found in \cite{tong2016}.

The description above, in terms of spin-singlet states, is rather different from the original construction of Blok and Wen \cite{blok-wen}. They constructed their eponymous states as conformal blocks of the $SU(p)_k$ Wess-Zumino-Witten (WZW) model. In other words, the Blok-Wen states can be written in the form \eqn{eq:trial}. To specify the state, one must decide which primary operator ${\cal O}$ is associated to the electron. The primary operators of the $SU(p)_k$ WZW model are labeled by the representations of $SU(p)$, with the restriction that the corresponding Young diagram has no more than $k$ boxes in its upper row. The restriction that ${\cal O}$ form a simple current is more restrictive: it was  shown in \cite{SY1} that there are $p$ such simple currents, which we denote as ${\cal O}_n$ with $n=0,\ldots, p-1$. Here ${\cal O}_0=\mathbb{I}$ is the identity operator, while ${\cal O}_n$ transforms in the representation corresponding to the Young diagram 
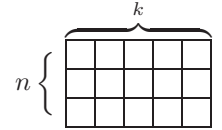
\be  n\left\{ \begin{array}{c} \\ \\ \end{array}\right. \!\!\!\overbrace{\raisebox{-3.1ex}{\yng(5,5,5)}}^{k} \ .
\nn\ee
In words, this is the $k$-fold symmetrization of the $n^\text{th}$ antisymmetric representation. The statement that these operators are simple currents means that they fuse only among themselves. Indeed, one finds that
the operator product expansion of ${\cal O}_n(z)$ with ${\cal O}_m(z')$ has the form
\begin{align}
		\mathcal{O}_n(z)&\mathcal{O}_m(z') \sim \nn \\
& \frac{1}{(z-z')^{h_{m} +  h_{n} -h_{m+n\ {\rm mod}\ p}}}\mathcal{O}_{m+n\ {\rm mod}\ p}(z') 
+ \dots\ ,
\nn\end{align}
where $h_m$ is the conformal weight of ${\cal O}_n$ and the ellipses denote terms which are regular as $z\to z'$.

The Blok-Wen states are built from correlation functions of the primary operator ${\cal O} = {\cal O}_1$,  transforming in the same $k^\text{th}$ symmetric representation of $SU(p)$ as the ground
states \eqref{eq:single-particle-states} of the matrix model for a single particle. 
The fact that this is the same representation as the particles in the matrix model is the first clue that the two states are the same. A full explanation of the equivalence can be found in \cite{tong2016}.

Given that the Blok-Wen states can be written in terms of conformal blocks \eqn{eq:trial}, we are in a position to apply Read's formula \eqn{read} to determine the Hall viscosity of this state. For this, we need the conformal weight $h_n$ of the operators ${\cal O}_n$. This is given by (see, for example, Section 3 of \cite{SY1}), 
\beq
	h_n= \frac{k}{2p}n(p-n)\ .
\eeq
Using this, together with the expression for the filling fraction \eqn{ffraction}, we find that the
guiding center contribution to the Hall viscosity is expected to be
\beq
	\eta_{\text{{\tiny{GC}}}} = \frac{\hbar}{2}\left( \frac{k+p}{2p} -\frac{1}{2} + \frac{k}{2p}(p-1)\right) \rho_0 =  \frac{\hbar k}{4}\rho_0\ .
\label{matchthis}\eeq
The apparent lack of dependence on the underlying $SU(p)$ symmetry is illusory: the density $\rho_0$ depends on both $k$ and $p$ as seen in \eqn{density}. However, the anisospin, defined by $\eta_{\ttt{GC}} = \frac{1}{2}\hbar \varsigma \rho_0$, is given by $\varsigma = k/2$ and is independent of $p$.  In the next 
section we show that the Hall viscosity of the matrix model coincides with the guiding center Hall viscosity \eqref{matchthis}
of the Blok-Wen states.

\section{Hall viscosity in the matrix model}\label{mmhvsec}

The computation of the Hall viscosity from the quantum Hall matrix model with $p=1$ was performed recently in \cite{lapa2018}. In many ways, the computation is straightforward: indeed, as we will explain below, one can extract the Hall viscosity from the angular momentum \eqn{l0}. The difficult part is to understand how to do this.

In general, the Hall viscosity of a quantum Hall state can be computed by studying the response of that state to 
time-dependent
area-preserving deformations (APDs). In the context of the matrix model, these APDs should be applied to the matrix 
coordinates $X^a$; it was shown in \cite{lapa2018} that the APDs are generated by the operators
\beq
	\mathsf{\Lambda}^{ab}= \frac{1}{4\ell_B^2}\sum_{A=0}^{N^2-1}\{x^a_A,x^b_A\}\ ,
\eeq
where $\{\cdot,\cdot\}$ denotes an anti-commutator and 
$x^a_A$ are the real scalar variables that we
 introduced in Eq.~\eqref{eq:parametrization} (and recall that these variables become Hermitian operators in the quantized
theory, with commutation relations given in Eq.~\eqref{eq:x-relations}). These operators obey the Lie algebra of the group 
$SL(2,\mathbb{R})$,
\beq
	[\mathsf{\Lambda}^{ab},\mathsf{\Lambda}^{cd}] = \frac{i}{2}\left(\ep^{bc}\mathsf{\Lambda}^{ad} + \ep^{bd}\mathsf{\Lambda}^{ac} + \ep^{ac}\mathsf{\Lambda}^{bd}+\ep^{ad}\mathsf{\Lambda}^{bc}  \right)\ .
\eeq
Finite APDs are specified by a symmetric matrix $\al$ 
of ``strain parameters," with components $\al_{ab}=\al_{ba}$. 
For the case of a spatially uniform APD the components $\al_{ab}$ are 
constants. The finite APD specified by $\al$ is implemented by the unitary operator 
$U(\al)=e^{i\al_{ab}\mathsf{\Lambda}^{ab}}$, and this operator acts on the matrix elements of the noncommutative 
coordinates as
\beq
	U(\al) {(X^a)^j}_k U(\al)^{\dg}= {(X^a)^j}_k + \ep^{ab}\al_{bc}{(X^c)^j}_k+\cdots \ .
\eeq 
These operators perform the same deformation on every matrix element ${(X^a)^j}_k$ of $X^a$. Therefore,
we can identify the operators $\mathsf{\Lambda}^{ab}$ as the correct generators of APDs of the full noncommutative (i.e.,
matrix-valued) coordinates $X^a$. For more details on this identification, and for a comparison with the APD generators
of the electron coordinates in the quantum Hall problem, we refer the reader to Ref.~\onlinecite{lapa2018}.
 
The Hall viscosity tensor for the matrix model is defined as follows. First, we define a generalized force $F^{ab}$ 
associated with an APD of the matrix model Hamiltonian. Let 
$H_{\text{{\tiny{MM}}}}(\al)= U(\al)H_{\text{{\tiny{MM}}}}U(\al)^{\dg}$ be the deformed 
Hamiltonian. Then the generalized force is $F^{ab}= -{\pd H_{\text{{\tiny{MM}}}}(\al)}/{\pd \al_{ab}}\big|_{\al=0}$.
Next, we subject the system to a \emph{time-dependent} APD specified by a symmetric matrix $\al(t)$, and we study the 
expectation value of the generalized force $F^{ab}$ in the ground state $|\psi(t)\ran$ of the time-dependent Schrodinger 
equation with Hamiltonian $H_{\text{{\tiny{MM}}}}(\al(t))$. More precisely, we study the expectation value
$\lan \psi(t)| U(\al(t))F^{ab}U(\al(t))^{\dg} |\psi(t)\ran$, where the operator $U(\al(t))F^{ab}U(\al(t))^{\dg}$ is equal to
the force $F^{ab}$ expressed in terms of the \emph{deformed} coordinates (i.e. we view the APD as 
an active transformation of the system~\cite{bradlyn2012}). In addition, we assume that the strain parameters vanish at the initial time $t_0$,
$\al_{ab}(t_0)= 0$, so that $|\psi(t_0)\ran$ is equal to the ground state $|\psi_0\ran$ of the undeformed Hamiltonian 
$H_{\text{{\tiny{MM}}}}$. Under these assumptions the expectation value of $U(\al(t))F^{ab}U(\al(t))^{\dg}$ has an 
expansion in terms of time derivatives of $\al(t)$ which takes the form
\beq\begin{split}
\lan \psi(t)| U(\al(t))F^{ab}U(\al(t))^{\dg} |\psi(t)\ran & \\
 &\hspace{-9em} = \lan \psi_0| F^{ab}|\psi_0\ran + \Gamma^{abcd}\dot{\al}_{cd}(t) + \dots\ ,
\end{split}\eeq
where $\Gamma^{abcd}$ is a tensor which is independent of $\al(t),$ and the ellipses denote terms which are higher
order in time derivatives of $\al(t)$. The Hall viscosity tensor $\eta^{abcd}_{\text{{\tiny{MM}}}}$ of the matrix model
is then equal to 
\beq
	\eta^{abcd}_{\text{{\tiny{MM}}}}= \frac{\Gamma^{abcd}}{\mathcal{A}}\ ,
\eeq
where $\mathcal{A}$ denotes the area of the droplet of quantum Hall fluid represented by the ground state of the matrix
model. The physical meaning of the tensor $\eta^{abcd}_{\text{{\tiny{MM}}}}$ is that it determines the linear response of the
``generalized stress" $\frac{1}{\mathcal{A}} {U(\al(t))F^{ab}U(\al(t))^{\dg}}$ to the ``rate of strain" $\dot{\al}_{cd}(t)$.

\vskip 4mm
When the system has rotational symmetry the four index tensor $\eta^{abcd}_{\text{{\tiny{MM}}}}$ has only one
independent coefficient, which we denote by $\eta_{\text{{\tiny{MM}}}}$. In \cite{lapa2018} it was shown that
this coefficient can be expressed in terms of the total angular momentum $L_0$ of the ground state $|\psi_0\ran$  as
\beq
	\eta_{\text{{\tiny{MM}}}}= \frac{1}{2}\frac{L_0}{\mathcal{A}}\ .
\eeq
At this stage, there is a slight complication: the  Hall viscosity coefficient computed from $L_0$ given in \eqn{l0} diverges as 
$N\to \infty$. This arises because of the term in $L_0$ proportional to $N^2$ (recall from Eq.~\eqref{eq:MM-area} that 
$\mathcal{A}$ is proportional to $N$ for the matrix model).
To handle this divergent term we take the same approach as in \cite{lapa2018} and
use the regularization procedure of Park and Haldane~\cite{park-haldane}. Park and Haldane noted that the term in 
$L_0$ proportional to $N^2$ represents the orbital angular momentum of the quantum Hall fluid, and they proposed a 
regularization procedure in which this term is simply subtracted to define the regularized Hall viscosity. Their rationale for this
was that, after this subtraction, the regularized Hall viscosity coefficient is proportional to the 
spin angular momentum per particle, which is identified with the coefficient of the
term in $L_0$ proportional to $N$, in  accordance with the physical interpretation of the Hall viscosity coefficient given in \cite{read2009,read-rezayi}. To this end we define
\beq
	L_{orb}= -\hbar\left(\frac{k+p}{p}\right)\frac{N^2}{2}\ ,
\eeq
and then the regularized Hall viscosity coefficient for the matrix model is given by 
\beq
	\eta_{\text{{\tiny{MM}},reg}} = \frac{L_0-L_{orb}}{2\mathcal{A}} = \frac{\hbar k}{4}\rho_0 \ . \label{eq:MMHV}
\eeq
The result exactly matches the \emph{guiding center} Hall viscosity of the Blok-Wen states \eqn{matchthis}, as calculated using 
Read's formula \eqref{read}, and the relevant input from the boundary CFT. 
Therefore, as in \cite{lapa2018}, we find that the matrix model captures the guiding center 
contribution to the Hall viscosity, but lacks the trivial Landau orbit contribution $\hbar\rho_0/4$. 
As in \cite{lapa2018}, the reason for this can
be traced back to the fact that the matrix model describes the corresponding quantum Hall state after projection into the lowest
Landau level, and all information about the Landau orbit degrees of freedom is lost in this projection.



\section{Conclusion}

We have extended the Hall viscosity calculation of Ref.~\onlinecite{lapa2018} to the quantum Hall matrix model for
the Blok-Wen series of non-Abelian fractional quantum Hall states~\cite{tong2016,MM-WZW}. Our result for the Hall 
viscosity of the non-Abelian matrix model is in complete agreement with the expected result for the Blok-Wen states as
derived from Read's general formula for the 
Hall viscosity of quantum Hall trial states constructed from conformal blocks in a CFT~\cite{read2009}. More precisely, we find
that the matrix model captures the \emph{guiding center} part of the Hall viscosity of the Blok-Wen states, but lacks the
trivial Landau orbit contribution, and this can be traced back to the fact that the matrix model describes the quantum Hall
states after projection into the lowest Landau level.

A reachable goal for future work would be to study the density profile near the boundary of the quantum Hall droplet
described by the matrix model. Previous studies of the boundary of Laughlin, and other, fractional quantum Hall states
have revealed the presence of a boundary \emph{double layer} accompanied by an electric dipole moment per unit
length with a quantised value~\cite{wiegmann2012}. Remarkably, this edge dipole moment has also been shown to be directly related
to the Hall viscosity in the bulk~\cite{park-haldane,CFTW} (see also Appendix C of 
Ref.~\onlinecite{read-rezayi}). It would be interesting to investigate this boundary layer physics, and the relation to the bulk
Hall viscosity, in the context of the exactly solvable quantum Hall matrix models.

\acknowledgements

We thank Andrey Gromov for useful discussions. M.F.L and T.L.H acknowledge support from the US National
Science Foundation under grant DMR 1351895-CAR, as well as the support of the Institute
for Condensed Matter Theory at the University of Illinois at Urbana-Champaign.   D.T and C.P.T  are supported  by the STFC consolidated grant ST/P000681/1. D.T is a 
Wolfson Royal Society Research Merit Award holder.  
C.P.T is supported by a Junior Research Fellowship at Gonville \& Caius College, Cambridge.


%

\end{document}